\documentclass[twocolumn,english,aps,prb,superscriptaddress,showkeys,pdftex,reprint]{revtex4-1}
\usepackage[latin9]{inputenc}
\usepackage{verbatim}
\usepackage{amsmath}
\usepackage{amssymb}
\usepackage{graphicx}

\makeatletter

\providecommand{\tabularnewline}{\\}

\usepackage {url}

\makeatother

\usepackage{babel}
\begin{document}

\title{Tunable third-harmonic generation from polaritons in the ultrastrong
coupling regime}

\author{Fábio Barachati}
\email{fabio-souza.barachati@polymtl.ca}

\selectlanguage{english}%

\affiliation{Department of Engineering Physics, École Polytechnique de Montréal,
Montréal H3C 3A7, QC, Canada}

\author{Janos Simon}

\affiliation{School of Chemistry and Biochemistry and Center for Organic Photonics
and Electronics, Georgia Institute of Technology, Atlanta, GA 30332-0400,
USA}

\author{Yulia A. Getmanenko}

\affiliation{School of Chemistry and Biochemistry and Center for Organic Photonics
and Electronics, Georgia Institute of Technology, Atlanta, GA 30332-0400,
USA}

\author{Stephen Barlow}

\affiliation{School of Chemistry and Biochemistry and Center for Organic Photonics
and Electronics, Georgia Institute of Technology, Atlanta, GA 30332-0400,
USA}

\author{Seth R. Marder}

\affiliation{School of Chemistry and Biochemistry and Center for Organic Photonics
and Electronics, Georgia Institute of Technology, Atlanta, GA 30332-0400,
USA}

\author{Stéphane Kéna-Cohen}
\email{s.kena-cohen@polymtl.ca}

\selectlanguage{english}%

\affiliation{Department of Engineering Physics, École Polytechnique de Montréal,
Montréal H3C 3A7, QC, Canada}

\date{\today}
\begin{abstract}
Strong inter-particle interactions between polaritons have traditionally
stemmed from their exciton component. In this work, we impart a strong
photonic nonlinearity to a polaritonic mode by embedding a nonlinear
polymethine dye within a high-Q all-metal microcavity. We demonstrate
nonlinear microcavities operating in the ultrastrong coupling regime
with a normalized coupling ratio of 62\%, the highest reported to
date. When pumping the lower polariton branch, we observe tunable
third-harmonic generation spanning the entire visible spectrum, with
internal conversion enhancements more than three orders of magnitude
larger than in bare films. Transfer matrix calculations indicate that
the observed enhancements are consistent with the enhanced pump electric
field. 
\end{abstract}

\keywords{Third-harmonic generation, ultrastrong coupling, exciton-polaritons,
microcavities, organic semiconductors, cavity QED.}
\maketitle

\section{Introduction}

Semiconductor microcavities are natural platforms for the study of
light-matter interaction due to their ability to confine cavity photons
and semiconductor excitons to the same region of space.\cite{RevModPhys.85.299}
If their interaction is strong enough, hybrid light-matter states
are formed, called upper (UP) and lower (LP) polaritons.\cite{PhysRevLett.69.3314,lidzey1998strong}
They are separated in energy by the vacuum Rabi splitting ($\Omega_{R}$),
which is related to the coupling strength. Much of the interest in
the strong light-matter coupling regime stems from the broad range
of nonlinear phenomena that can be observed using polaritons, such
as parametric scattering, amplification, bistability, soliton formation
and superfluidity.\cite{PSSB:PSSB200560960,sanvitto2016road,RevModPhys.85.299}
Most of these phenomena have, however, been restricted to inorganic
microcavities at low-temperatures. There, the intrinsic nonlinearity
responsible for pairwise scattering, similar to an optical $\chi^{(3)}$,
is strong due to the delocalized nature of Wannier-Mott excitons.
The large binding energy of organic Frenkel excitons can in principle
allow for such phenomena to be observed at room-temperature, but the
exciton-exciton nonlinearities inherent to Frenkel excitons tend to
be much weaker than in inorganics. To date, only a handful of nonlinear
processes have been observed using organic polaritons, \cite{PhysRevB.69.235330,PhysRevB.74.113312,kena2010room,daskalakis2014nonlinear,plumhof2014room}
including room-temperature superfluidity.\cite{lerario2016room}

\begin{figure}[ht]
\includegraphics[width=0.45\textwidth]{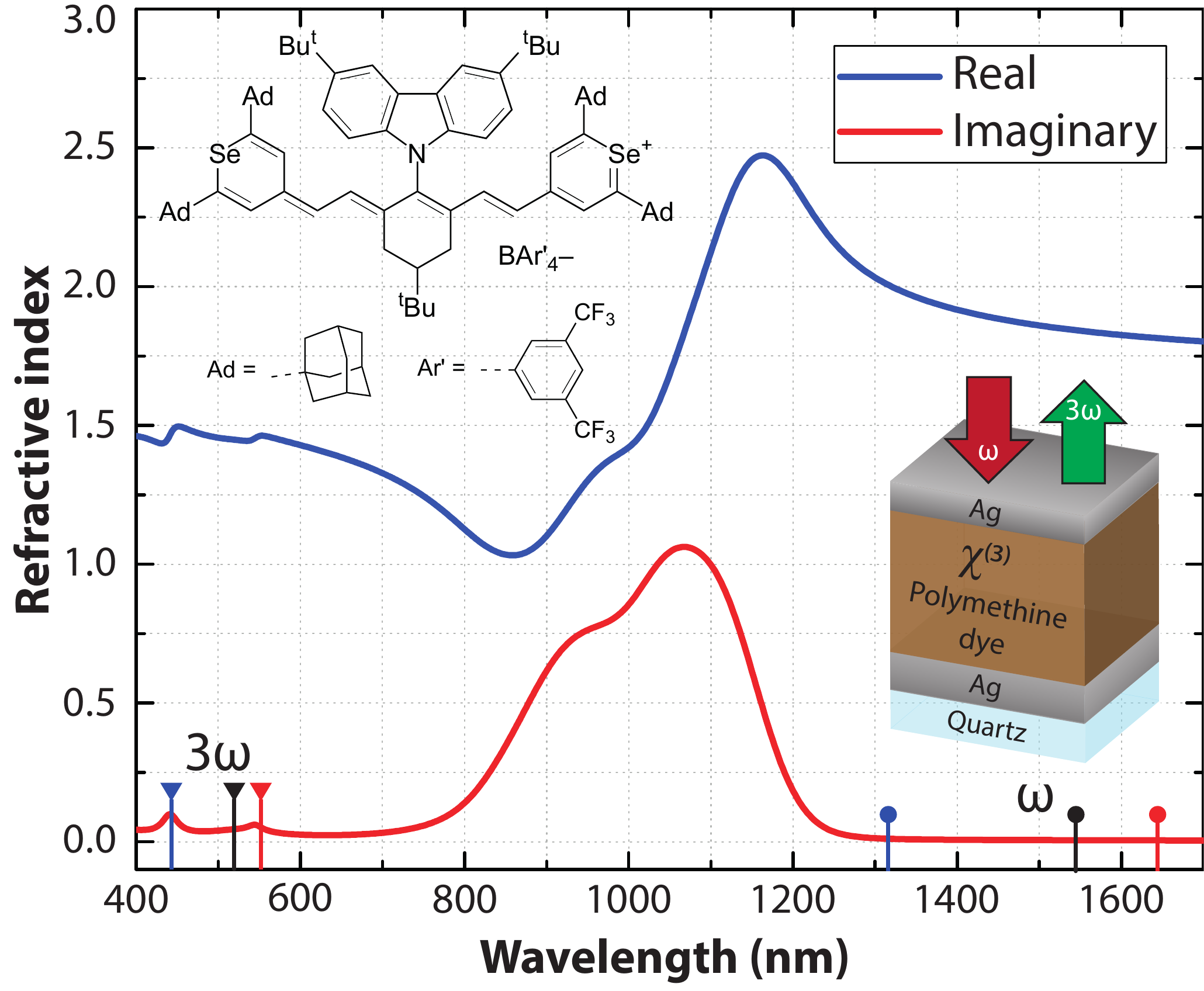} \caption{Real (blue) and imaginary (red) refractive index of neat polymethine.
The imaginary part has a peak value of 1.06 at 1067 nm. Round (triangular)
markers indicate the pump (THG) wavelengths where power dependence
measurements were performed. The inset shows (top) the chemical structure
of the polymethine dye and counterion and (bottom) a schematic of
the microcavity structure.}
\label{FigNk} 
\end{figure}

\begin{figure*}[t]
\includegraphics[width=1\textwidth]{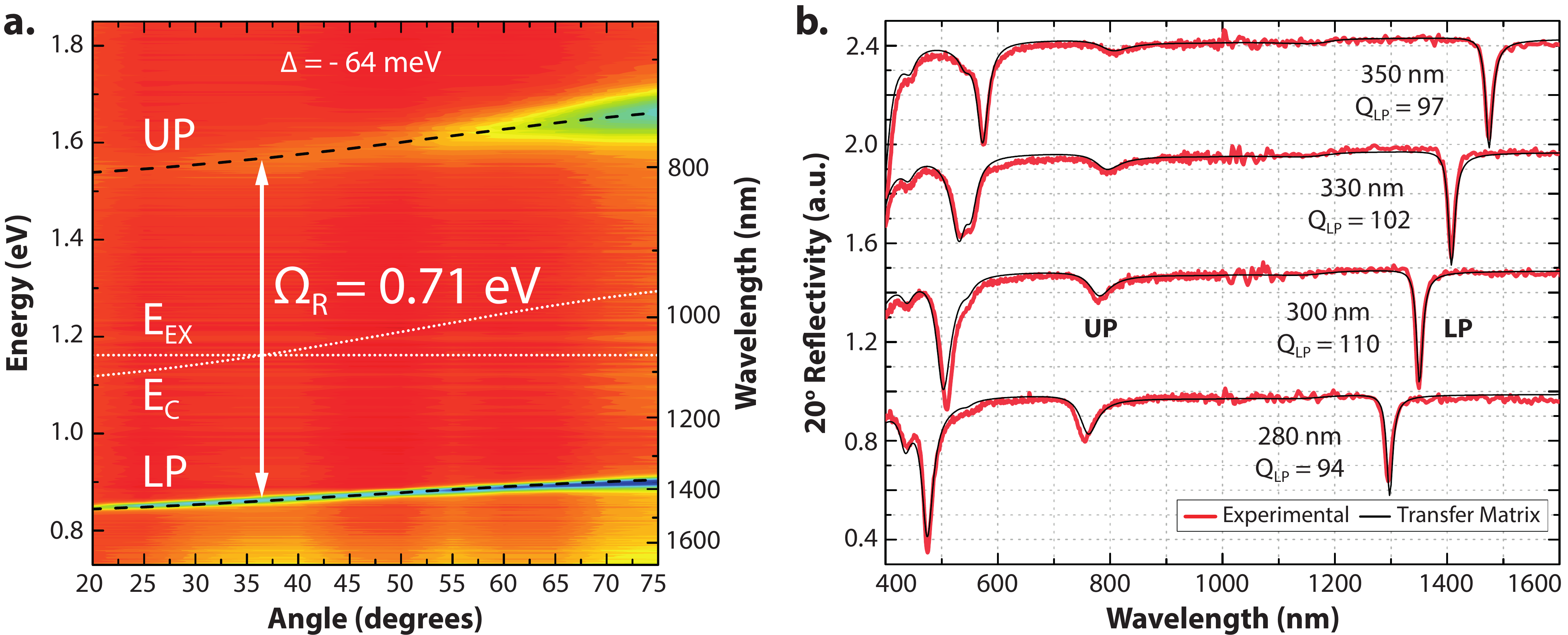} \caption{(a) False-color angle-dependent TM-polarized reflectivity spectrum
of a 350 nm-thick microcavity. The dashed lines show the least-squares
fit to the full Hopfield Hamiltonian, yielding a Rabi splitting of
$\Omega_{R}=0.707$ eV and a cavity energy at normal incidence of
$E_{C}=1.098$ eV. The detuning is $\Delta=E_{C}-E_{EX}=-64$ meV.
(b) Measured (red) and calculated (black) TM-polarized $20^{o}$ reflectivity
spectra for regions of different thickness present on the same sample,
as well as the experimental LP quality factors. The dip below 600
nm is the UP that originates from coupling to the second-order cavity
mode.}
\label{FigRefl} 
\end{figure*}

In the limit where the exciton-photon coupling is increased to a significant
fraction of the uncoupled exciton energy ($\Omega_{R}\sim E_{EX}$),
the system enters the so-called ultrastrong coupling regime (USC).
In this regime, the presence of non-negligible antiresonant light-matter
coupling terms leads to modifications of both excited and ground state
properties, a subject which is currently under intense investigation.\cite{PhysRevB.72.115303,PhysRevLett.112.016401,PhysRevLett.116.238301}
Demonstrations of USC in organic microcavities\cite{PhysRevLett.106.196405,ADOM:ADOM201300256,doi:10.1021/ph500266d}
have triggered many interesting studies about its effects on observable
material properties,\cite{ANIE:ANIE201107033,PhysRevX.5.041022,orgiu2015conductivity}
with many open questions remaining.

In this paper, we demonstrate organic microcavities operating in the
USC regime with a Rabi splitting corresponding to a record 62\% of
the uncoupled exciton energy. The microcavities are composed of a
film of a bis(selenopyrylium)-terminated heptamethine dye (see Fig.
\ref{FigNk}). Dyes of this type exhibit extremely large magnitudes
of the molecular third-order polarizability in solution.\cite{Hales1485}
The dye used in this study is one of many developed in which bulky
substituents on both the chalcogenopyrylium end groups and on the
polymethine chain, along with a large counterion, are used to disrupt
intermolecular interactions such that the solution linear and nonlinear
properties are largely preserved in high chromophore-density films.
\cite{C4MH00068D} We show that the resulting large magnitude of the
third-order susceptibility $\chi^{(3)}$ contributes to enhancing
polariton-polariton interactions via the photonic component. This
is used to demonstrate tunable third-harmonic generation (THG) spanning
the entire visible spectrum upon resonant excitation of the LP mode.
When compared to bare films, the fabricated microcavities show conversion
efficiency enhancements of over two orders of magnitude and even larger
internal enhancements. The structure forms a versatile platform for
the study of nonlinear phenomena in the USC regime. 

\section{\label{ResultsAndDiscussion}Results and discussion}

To characterize the linear optical properties of the polymethine dye,
neat films were prepared by spin-coating from 5-20 mg/ml dichloromethane
solutions. The refractive index obtained using ellipsometry is shown
in Fig.~\ref{FigNk}. The imaginary component shows a strong exciton
absorption maximum at 1067 nm ($E_{EX}=1.162$ eV) and a vibronic
shoulder. Note in particular the low losses in the near-infrared part
of the spectrum.

The fabricated microcavities are composed of a neat polymethine layer
surrounded by silver mirrors and are shown schematically in the bottom
inset of Fig.~\ref{FigNk}. To improve the wetting and optical properties
of the back mirror, cleaned quartz substrates were first functionalized
with a monolayer of (3-mercaptopropyl)-trimethoxysilane. \cite{doi:10.1063/1.4963262}
Then, a 75 nm-thick silver mirror was grown by thermal evaporation
at a base pressure of $\sim$$10^{-7}$ mBar. After spin-coating the
polymethine film, the structure was capped with a 35 nm top mirror.
All of the measurements were performed at room temperature. 

\begin{figure*}[t]
\includegraphics[width=1\textwidth]{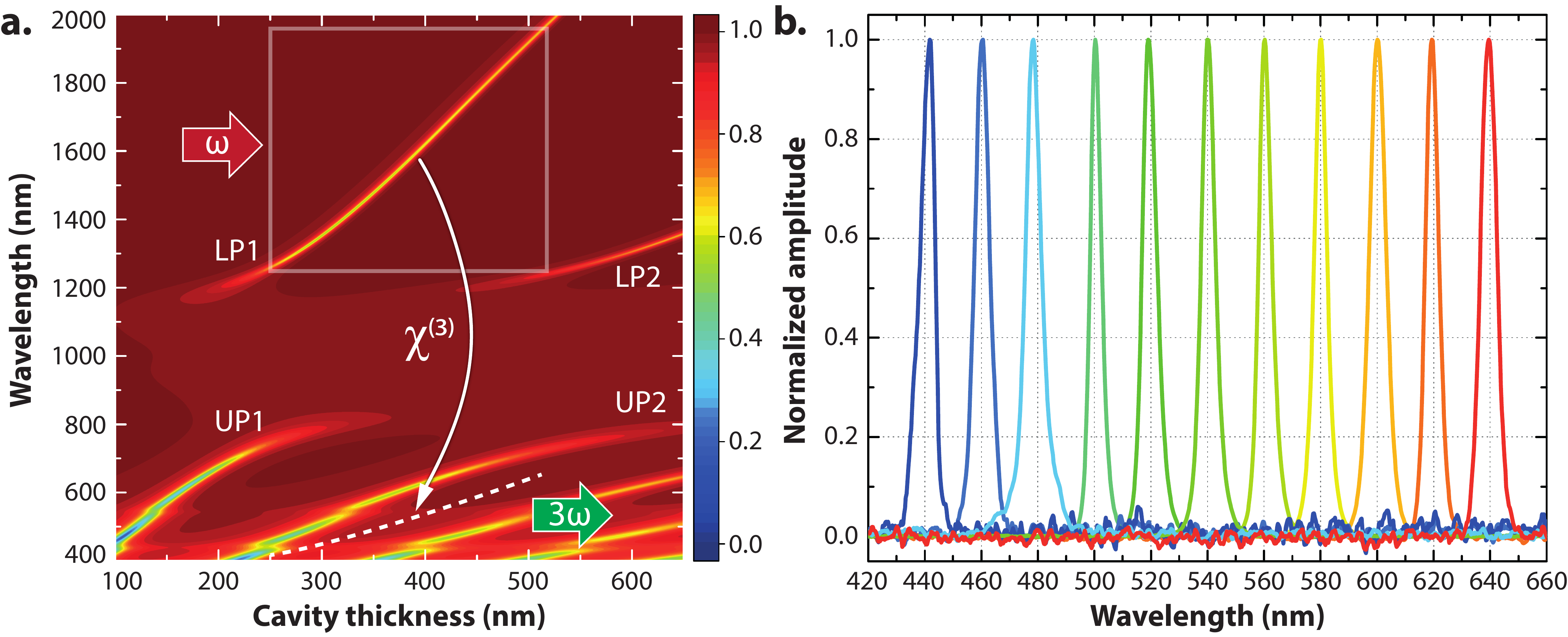} \caption{(a) Calculated microcavity reflectivity spectra at normal incidence
for increasing values of polymethine thickness. The four polaritonic
modes indicated by LP1/UP1 and LP2/UP2 arise due to the strong coupling
of the exciton transition to the first- and second-order cavity modes,
respectively. The top box indicates the range of LP spectral positions
covered by the broad infrared pump (1250-1950 nm). The corresponding
third-harmonic wavelengths are indicated by a dashed line. (b) Experimental
normalized THG spectra obtained by laterally translating the same
microcavity sample.}
\label{FigTHGRS} 
\end{figure*}

Figure~\ref{FigRefl} (a) shows the measured angle-resolved TM-polarized
reflectivity from a 350 nm thick microcavity, which corresponds to
nearly zero detuning between the exciton and photon energies at normal
incidence. The dashed lines correspond to a least-squares fit to the
full Hopfield Hamiltonian (see Methods),\cite{ADOM:ADOM201300256}
yielding a Rabi splitting of $\Omega_{R}=0.707$ eV. A slightly larger
value of $\Omega_{R}=0.719$ eV is obtained for the TE-polarized reflectivity
spectrum and the remaining fit parameters are summarized in Table~\ref{tbl:fit2}.

These values correspond to 60-62\% of the uncoupled exciton energy,
slightly exceeding the largest normalized coupling ratio ($\Omega_{R}/E_{EX}$)
reported to date of 60\% in organic microcavities.\cite{doi:10.1021/ph500266d}
The large ratio observed is a consequence of the high oscillator strength
of the dye and the large number density of molecules resulting from
the use of a neat film, both evidenced by the strong absorption band
shown in Fig.~\ref{FigNk}. In addition, the use of metallic mirrors
instead of dielectric ones leads to a reduced photonic mode volume.
These two factors contribute to the increased $\Omega_{R}$. Meanwhile,
the near-infrared transition energy leads to a smaller $E_{EX}$ than
in previous reports. Note that for larger cavity thicknesses, the
coupling ratio can exceed 90\% in this structure.

We find that the experimental LP quality factors can exceed 100, which
is considerably higher than typical values (below 30) obtained for
all-metal microcavities.\cite{doi:10.1063/1.1517714,PhysRevB.77.073205,PhysRevLett.106.196405}
This is a consequence of the material's low linear losses and the
high reflectivity of silver in the near-infrared part of the spectrum.
This is helpful for increasing the efficiency of nonlinear processes
or, in some cases, lowering their thresholds.\cite{PhysRevB.62.R4825,Rodriguez:07}
Despite the inhomogeneously broadened absorption of the dye, the LP
reflectivity spectra showed Lorentzian lineshapes characteristic of
homogeneous broadening, a consequence of the large Rabi splitting
compared to the inhomogeneous linewidth.\cite{PhysRevA.53.2711}

The LP resonance position can be readily tuned via changes in angle
of incidence or sample thickness. Our use of high concentration solutions
and of a high vapor pressure solvent naturally leads to large-scale
thickness gradients (hundreds of microns) over the sample surface.
This allows for multiple cavity thicknesses to be probed using a single
sample. These can be easily identified experimentally due to changes
in the surface color caused by the changing position of the UP branch.
Figure~\ref{FigRefl} (b) shows a collection of measured (red) and
calculated (black) TM-polarized $20^{o}$ reflectivity spectra taken
at different locations. The corresponding sample thicknesses are indicated
below the traces.

Third-harmonic generation was first studied by exciting the sample
with a broad IR pump spanning 1250-1950 nm. In the experiment, a single microscope
objective is used to focus the IR pump on the sample surface and to
collect the reflected third-harmonic (see Methods). Figure \ref{FigTHGRS}
(a) shows a calculated microcavity reflectivity map, where the spectral
positions of the polariton modes are obtained for increasing values
of cavity thickness. The first four polaritonic modes indicated by
LP1/UP1 and LP2/UP2 correspond to strong coupling to the first- and
second-order cavity modes, respectively. The broad pump is spectrally
filtered by the thickness-dependent LP resonance position, as shown
in the top box in Fig.~\ref{FigTHGRS} (a). The resonant component
is coupled into the microcavity and interacts with the high $\chi^{(3)}$
material to generate the third-harmonic signal. The THG wavelengths
corresponding to the first LP branch are shown as a dashed line in
the bottom of Fig.~\ref{FigTHGRS} (a). Figure~\ref{FigTHGRS} (b)
shows a series of normalized THG spectra taken by laterally translating
the sample and illustrates the tunability of the THG process at normal
incidence. Note that the third-harmonic is not resonant with higher
order modes and only light that can leak out due to the finite transmission
of the mirrors is observed. Nevertheless, the generated harmonic signals
were easily visible with the naked eye through the microscope optics.

\begin{figure}[t]
\includegraphics[width=0.45\textwidth]{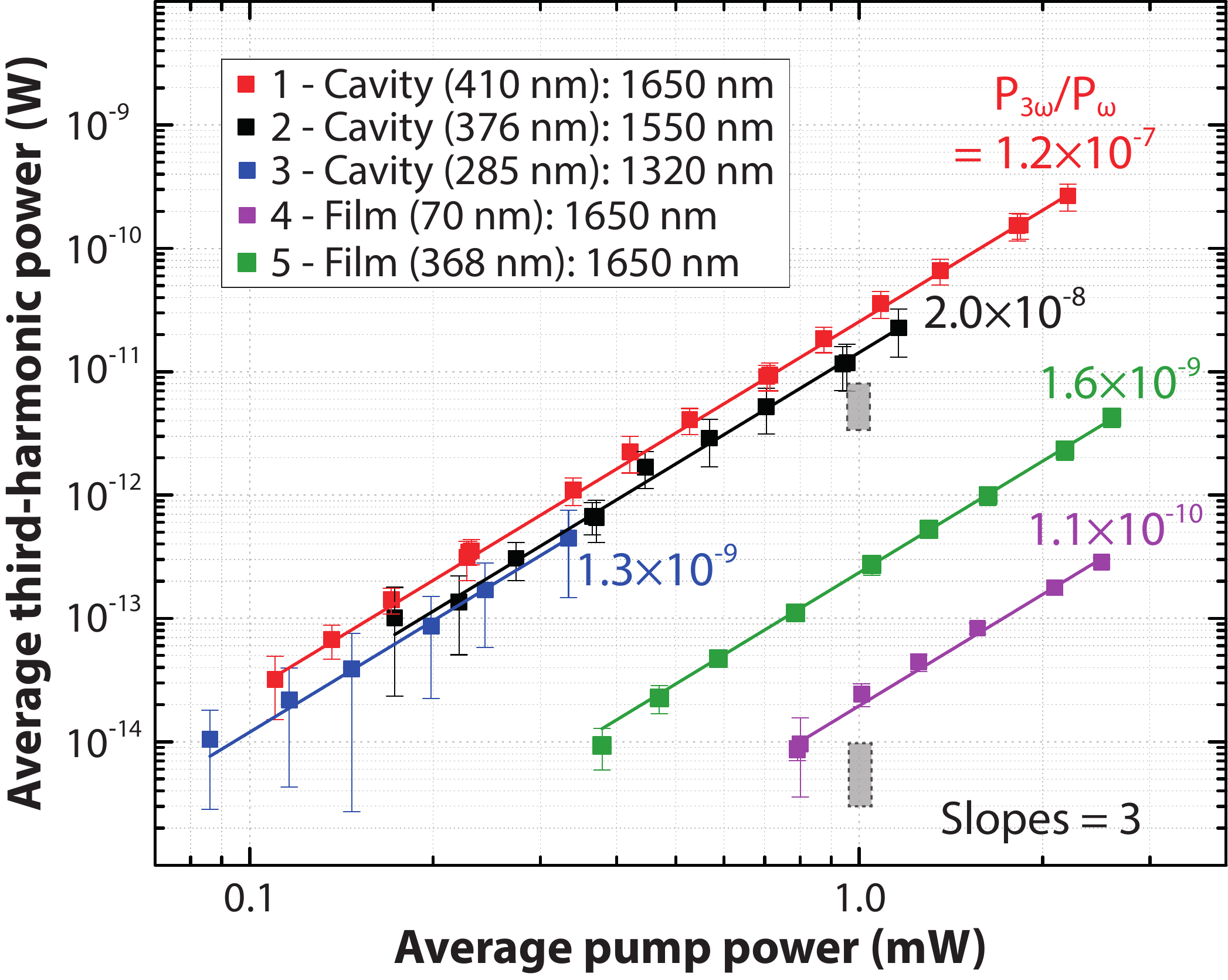} \caption{Measured third-harmonic powers (squares) as a function of pump power.
Red, black and blue data sets correspond to microcavities pumped at
1650 nm, 1550 nm and 1320 nm, respectively. Bare polymethine films
of 70 nm and 368 nm were investigated at 1650 nm and the results are
shown in purple and green, respectively. The solid lines indicate
a cubic power fit to each data set. The values at the end of each
curve correspond to their peak conversion efficiencies, defined as
$P_{3\omega}/P_{\omega}$. The top (bottom) gray boxes indicate the
range of calculated reflected THG powers for the microcavity (bare
films) shown in Fig.~\ref{FigTHGmap}. The pump spot had a FWHM of
$1.5~\mu$m.}
\label{FigPower} 
\end{figure}

To investigate the THG power dependence as a function of wavelength,
the broad IR pump was spectrally filtered using 12 nm bandpass filters,
which is slightly narrower than the mean LP linewidth of 14 nm obtained
from Fig.~\ref{FigRefl} (b). The pump and corresponding THG wavelengths
are indicated in Fig.~\ref{FigNk} by round and triangular markers,
respectively. Figure~\ref{FigPower} shows the results obtained for
the fabricated microcavities (curves 1-3). We observe a doubling of
the conversion efficiency when changing from 1320 nm to 1650 nm excitation.
This corresponds to an increased THG efficiency for polaritonic modes
that are more photonic in nature, as highlighted in Fig. \ref{FigTHGmap}
(a). When pumping at 1320 nm, which corresponds to modes with a strong
exciton component, we observed a fast irreversible decay of the THG
signal due to sample damage \textemdash{} a clear signature of the
mode matter content. Note, in contrast, that the linear absorption
of the bare film is negligible at this wavelength. For longer wavelengths,
no decrease in THG powers was observed for up to an hour of measurement. 

The values at the end of each curve correspond to the peak power conversion
efficiencies, defined as $P_{3\omega}/P_{\omega}$. In all cases,
the THG wavelength was not resonant with the higher order upper polariton
modes and was not efficiently extracted from the microcavity. Nonetheless,
the high experimental conversion efficiencies for the microcavities
were comparable to those of other systems that required patterning
or considerably thicker films.\cite{doi:10.1021/nl503029j,doi:10.1063/1.1771809,doi:10.1021/acs.nanolett.6b01958}

For comparison, THG from bare polymethine films of different thicknesses
are shown in Fig.~\ref{FigPower} (curves 4-5). The 368 nm-thick
film (curve 5) is of comparable thickness to the microcavities and
allows for an accurate estimate of the conversion enhancements. Table~\ref{tbl:fit}
shows the fit coefficients $C$ obtained by fitting each data set
with a cubic power dependence of the form $P_{3\omega}=C\times{P_{\omega}}^{3}$.
The last column shows the conversion enhancements with respect to
the 368 nm-thick film. At the wavelength of highest conversion efficiency
(1650 nm), the fabricated microcavities show raw (external) THG enhancement
factors of $108\pm1.7$ compared to the bare film. As shown in Fig.
\ref{FigTHGmap} (a), the internal efficiency at this wavelength is
approximately 50 times higher than this value due to the low fraction
of THG that is out-coupled from the cavity.

\begin{table}[ht]
\caption{Experimental cubic fit coefficients for data sets 1-5 shown in Fig.~\ref{FigPower}
and the THG conversion enhancement with respect to the 368 nm-thick
bare film.}
\label{tbl:fit} %
\begin{tabular}{ccc}
\hline 
\#  & $C$  & THG Enhancement \tabularnewline
\hline 
1  & $(254.2\pm2.1)\times10^{-4}$  & $108.0\pm1.7$ \tabularnewline
2  & $(142.9\pm2.7)\times10^{-4}$  & $60.7\pm1.4$ \tabularnewline
3  & $(119.7\pm2.7)\times10^{-4}$  & $50.6\pm1.3$ \tabularnewline
4  & $(195.8\pm7.4)\times10^{-7}$  & \textendash{} \tabularnewline
5  & $(235.4\pm3.2)\times10^{-6}$  & \textendash{} \tabularnewline
\hline 
\end{tabular}
\end{table}

\section{\label{Simulations}Simulations}

\begin{figure*}[t]
\includegraphics[width=1\textwidth]{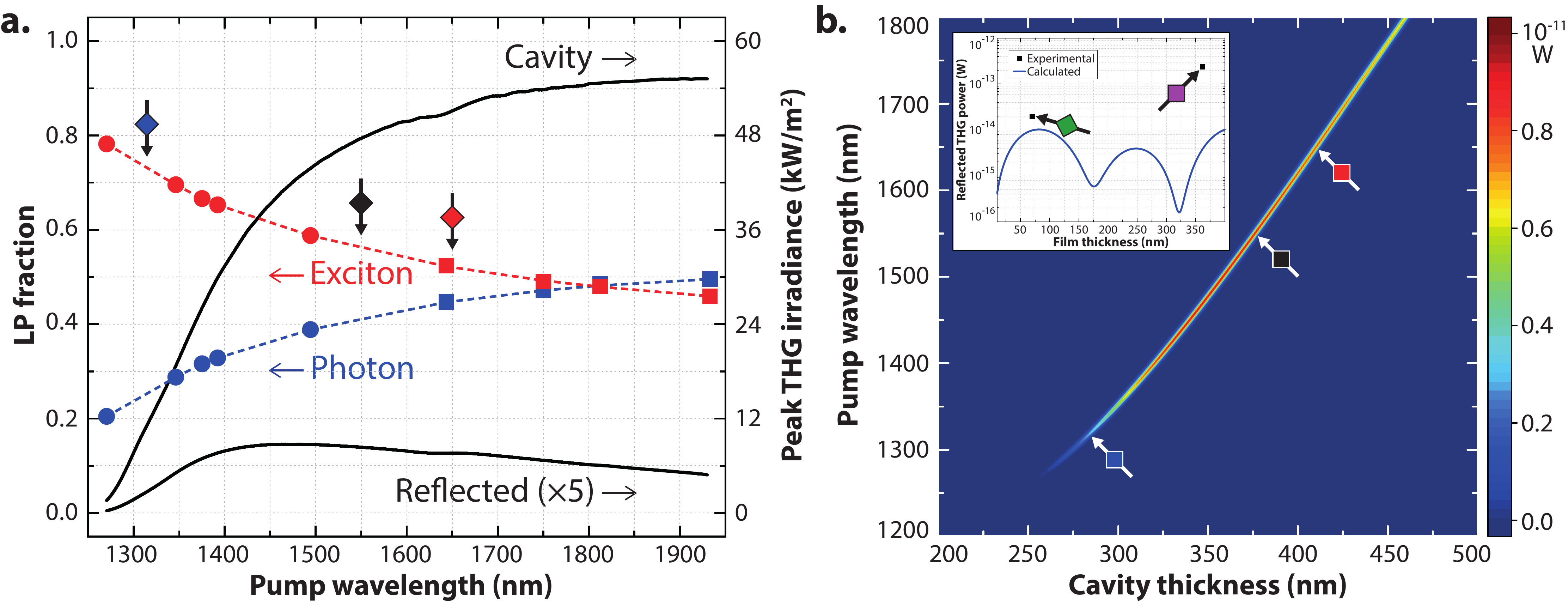} \caption{(a) Calculated exciton (red) and photon (blue) fractions for the first
LP branch from experimental (circles) and simulated (squares) reflectitivy
data. The solid black curves show the calculated THG irradiances inside
the cavity (top) and outside in reflection (bottom). (b) Calculated
reflected third-harmonic powers as a function of microcavity thickness
and pump wavelength for an average input power of 1 mW. THG is obtained
when the pump is resonant with the LP mode, maintaining the same shape
as in the boxed region in Fig.~\ref{FigTHGRS}. Inset: same calculation
for a thin film on quartz pumped at 1650 nm. For the calculations,
$\chi^{(3)}=-5.1\times10^{-19}$ esu ($-7.12\times10^{-19}~\mathrm{m}^{2}/\mathrm{V}^{2}$),
THG and pump spots with a FWHM of $1.5~\mu$m were used. The colored
arrows in a,b indicate the experimental conditions for the measurements
in Fig.~\ref{FigPower}. }
\label{FigTHGmap} 
\end{figure*}

To correlate our results with the LP exciton and photon fractions
at normal incidence, Fig. \ref{FigTHGmap} (a) shows the polariton
modal content calculated using the full Hopfield Hamiltonian from
experimental (circles) and simulated (squares) angle-resolved reflectivity
spectra. Note that in the USC regime, the contribution from the squared
electromagnetic vector potential leads to a blueshift of the bare
cavity photon energy and that the modal content of both components
is no longer equal at zero detuning, which corresponds to 1432 nm
in Fig. \ref{FigTHGmap} (a).\cite{PhysRevB.72.115303} We find that
the LP branch varies from 73\% to 52\% exciton content over the range
of thicknesses used in Fig.~\ref{FigPower} (indicated by colored
arrows). 

To determine the origin of the enhancement, THG was investigated using
a modified transfer matrix approach, which includes nonlinear source
terms.\cite{Bethune:89,PhysRevE.82.036608,PhysRevA.92.033828} In
this way, the combined effects of multiple reflections and absorption
at both the fundamental and third-harmonic wavelengths are automatically
taken into account, as well as the resonant pump enhancement and non-resonant
THG extraction factors.

Figure~\ref{FigTHGmap} (b) shows a map of the calculated reflected
THG powers for the fabricated microcavities as a function of cavity
thickness and pump wavelength for an average input power of 1 mW.
As expected, the THG map resembles the reflectivity map shown in the
boxed region of Fig.~\ref{FigTHGRS} (a) because THG is only generated
when the pump is resonant with the LP branch. The colored arrows indicate
the experimental conditions of cavity/film thickness and pump wavelength
for the measurements in Fig.~\ref{FigPower}. The top gray box in
Fig.~\ref{FigPower} shows the range of calculated reflected THG
powers for the microcavities. The quantitative agreement with the
experimental values is remarkable, considering that the only parameter
not known with certainty is the third-order susceptibility $\chi^{(3)}(3\omega;\omega,\omega,\omega)$,
which was kept dispersionless and equal to the value $\chi^{(3)}(\omega;-\omega,\omega,\omega)$
measured using the z-scan technique.\cite{C4MH00068D}

A similar calculation for the bare films pumped at 1650 nm is shown
in the inset of Fig.~\ref{FigTHGmap} (b) and was found to be less
accurate for the 368 nm-thick film, with the experimental value closer
to the transmitted THG power (not shown) instead of the reflected
one. The bottom gray box in Fig.~\ref{FigPower} shows the range
of calculated reflected THG powers for the bare films. We believe
that the discrepancy stems mostly from the large angular spread of
wavevectors in the focussed pump beam, which is not accounted for
in our calculations. For microcavities, wavevectors away from the
resonance condition are naturally filtered out by the LP dispersion. 

The agreement for the thinner film and microcavities confirm that
the value of $\chi^{(3)}$ used in the calculations is a reasonable
estimate for the real one. Therefore, even when neglecting possible
resonant enhancements of $\chi^{(3)}$ due to the weak absorption
bands in the visible part of the spectrum, the observed THG enhancements
seem to be consistent with a purely (or at least mostly) photonic
effect, as opposed to a drastic modification of the material's nonlinear
properties.\cite{nanolett.6b02567} Still, as compared to bare films,
microcavities offer the advantage that they require smaller thicknesses
to generate the same harmonic power. This is a consequence of the
pump electric field enhancement, the reduced absorption of the generated
signal and the absence of a phase-matching requirement, with the conversion
efficiency depending instead on mode overlap, which can be tuned by
microcavity design.\cite{Burgess:09} 

Fig.~\ref{FigTHGmap} (a) also shows the calculated peak THG irradiances
inside the cavity as compared to the reflected THG. Both curves significantly
decrease towards shorter wavelengths because of the absorption band
peaked at 1067 nm. The internal THG first increases towards longer
wavelengths as losses in the mirrors are reduced for thicker cavities.
The decreasing transmission of the front silver mirror towards longer
wavelengths eventually leads to a decrease in both the internal and
reflected components, the latter already visible in Fig.~\ref{FigTHGmap}
(a). For this range of wavelengths, the internal THG enhancements
are up to 56 times higher than the ones observed in reflection. Further
calculations show that small modifications of the structure to improve
out-coupling, such as reducing the top mirror thickness or achieving
a doubly-resonant condition of the THG wavelength with an UP branch
can further increase the THG efficiency by more than an order of magnitude.

\section{\label{Conclusion}Conclusion}

We have fabricated organic microcavities containing a nonlinear dye
operating in the USC regime and possessing a normalized coupling ratio
of 62\%, slightly exceeding the highest value ever reported.\cite{doi:10.1021/ph500266d}
The combination of the material's low losses and the use of high reflectivity
silver mirrors led to LP quality factors much higher than conventional
all-metal microcavities. The polariton nonlinearity was exploited
to demonstrate efficient and tunable third-harmonic generation when
the cavity was excited resonantly with the LP branch. Although the
THG was not resonantly extracted from the microcavities, conversion
enhancements of up to two orders of magnitude were obtained in comparison
to bare films and transfer matrix calculations indicate that the these
can be understood as due to the increased pump electric field. We
anticipate that this demonstration of a strong organic polariton nonlinearity
 will allow for the observation of novel room-temperature nonlinear
effects so far restricted to low-temperature inorganic systems and
new phenomena unique to the USC regime.

\section{Methods}

\subsection{Linear optical characterization}

An variable-angle ellipsometer (J. A. Woollam Co., RC2 D+NIR) was
used to obtain the thicknesses and optical constants of the neat polymethine
films, the silver mirrors and the quartz substrate. Angle-resolved
reflectivity measurements were performed using the same instrument
with the help of focusing probes.

\subsection{Fitting of the reflectivity data}

The experimental Rabi splitting and mixing coefficients were extracted
from the angle-resolved reflectivity data by performing a least-squares
fit to the eigenvalue equation

\begin{equation}
\hat{{H_{q}}}\vec{v}{}_{j,q}=\omega_{j,q}\vec{v}{}_{j,q}.
\end{equation}

$\hat{{H_{q}}}$ is the full Hopfield Hamiltonian given by

\begin{equation}
\hat{{H_{q}}}=\begin{pmatrix}E_{ph}(q)+2D & V & 2D & V\\
V & E_{EX} & V & 0\\
-2D & -V & -E_{ph}(q)-2D & -V\\
-V & 0 & -V & -E_{EX}
\end{pmatrix},
\end{equation}

where $q=(\omega/c)\sin\theta$ is the in-plane wavevector, $\theta$
is the angle of incidence, $V=\Omega_{R}/2$ is the interaction energy
and $D=V^{2}/E_{EX}$ is the energy contribution of the squared electromagnetic
vector potential.\cite{ADOM:ADOM201300256,PhysRevB.72.115303} The
eigenvectors $\vec{v}{}_{j,q}=(w_{j,q},x_{j,q},y_{j,q},z_{j,q})^{T}$are
obtained from the polariton annihilation operators 
\begin{equation}
p_{j,q}=w_{j,q}a_{q}+x_{j,q}b_{q}+y_{j,q}a_{-q}^{\dagger}+z_{j,q}b_{-q}^{\dagger},
\end{equation}

where $j\in\{LP,UP\}$ and $a,b$ are photon and exciton annihilation
operators, respectively. They provide the Hopfield coefficients for
the photon ($|w_{j,q}|^{2}$) and exciton ($|x_{j,q}|^{2}$) fractions
as well as the anomalous coefficients ($|y_{j,q}|^{2},|z_{j,q}|^{2}$)
which only become comparable to the first ones when $\Omega_{R}\sim E_{EX}$. 

The eigenvalues $\omega_{j,q}=\{\pm\omega_{LP,q},\pm\omega_{UP,q}\}$
correspond to the polariton energies to be matched with the reflectivity
minima. The fit parameters were the interaction energy $V$, the normal-incidence
cavity energy $E_{C}$ and the cavity's effective refractive index
$n_{eff}$, the last two being related by 

\begin{equation}
E_{ph}(q)=\sqrt{\left(\frac{{\hbar cq}}{n_{eff}}\right)^{2}+E_{C}^{2}}.
\end{equation}

The fit parameters obtained for the TE- and TM-polarized spectra are
summarized in Table~\ref{tbl:fit2}. 
\begin{table}[h]
\caption{Full Hopfield Hamiltonian fit parameters obtained for the reflectivity
data of a 350 nm cavity.}
\label{tbl:fit2} %
\begin{tabular}{cccc}
\hline 
Pol.  & $\Omega_{R}$ (eV)  & $n_{eff}$  & $E_{C}$ (eV) \tabularnewline
\hline 
TE  & $0.719\pm0.004$  & $1.57\pm0.04$  & $1.09\pm0.01$ \tabularnewline
TM  & $0.707\pm0.002$  & $1.83\pm0.04$  & $1.10\pm0.01$ \tabularnewline
\hline 
\end{tabular}
\end{table}

\subsection{Nonlinear optical characterization}

Third-harmonic measurements were performed in reflection using an
Olympus IX-81 inverted microscope. The samples were mounted facing
an Olympus LUCPlanFl 40$\times$ 0.6 NA objective with the correction
collar set to zero. A supercontinuum laser (Fianium FemtoPower 1060,
40 MHz) was used as the infrared light source. The pulse duration
was measured with a streak camera (Hamamatsu C10910) and found to
be 48 ps. A different set of excitation (EX), beam splitter (BS) and
detection (DET) filters was used for each fundamental wavelength.
All filters were purchased from Thorlabs unless indicated otherwise.
1320 nm: (EX) LP1250, BP1320-12, SP1500 (Edmund Optics), 1 cm quartz
cuvette with toluene; (BS) BSW29R; (DET) SP950, BP430-20 (Chroma),
1 cm quartz cuvette with water. 1550 nm: (EX) LP1250, LP1500, BP1550-12,
1 cm quartz cuvette with toluene; (BS) BSW29R; (DET) SP1500. 1650
nm: (EX) LP1250, LP1500, BP1650-12; (BS) FF538-FDi01 (Semrock); (DET)
1 cm cuvette with toluene. The transmission of all detection filters
was measured and used to calibrate the measured powers.

The third-harmonic powers were measured with a cooled CCD camera (Princeton
Instruments, PIXIS 400B eXcelon). The calibration procedure was performed
with a 532 nm laser (supercontinuum followed by BP532-3, SP700, SP1500,
1 cm cuvette with toluene). The laser was focused on a 75-nm thick
silver mirror and the reflected light was measured at the side-port
of the microscope with a power meter (Thorlabs, S120VC/PM100D). Neutral-density
filters with measured transmission values at 532 nm were inserted
in the excitation path so that the powers at the side-port of the
microscope could be adjusted in the range of $10^{-13}-10^{-10}$
W.

For each power value, a series of images was acquired with the CCD
camera and the corresponding integrated counts on two adjacent 30$\times$30
pixel regions were obtained, the first one including the laser spot
and the second one only background light. The procedure was repeated
for different integration times and the detector response was obtained
by fitting the background-corrected counts over the known input power
to a fourth-degree polynomial in integration time.

The CCD quantum efficiency is taken to remain constant in the range
of THG wavelengths from 440 nm to 550 nm. The reflection collection
efficiency of the objective based on the detection solid angle was
estimated to be $\eta=1-\cos\left[\sin^{-1}(NA)\right]=0.2$. This
value only affects absolute powers, but not the enhancement factors.
\begin{acknowledgments}
The authors would like to thank Johannes Feist and Simone de Liberato
for helpful comments on this manuscript. FB and SKC acknowledge support
from the Natural Sciences and Engineering Research Council of Canada
Discovery Grant Program, the Canada Research Chair in Hybrid and Molecular
Photonics and the FQRNT PBEEE scholarship program. Work at GT was
supported by the Air Force Office of Scientific Research through the
COMAS MURI program (agreement no. FA9550-10-1-0558).
\end{acknowledgments}

\end{document}